# The digital laser

Sandile Ngcobo[1,2], Igor Litvin[2], Liesl Burger[2] and Andrew Forbes[1,2*]

[1]School of Physics, University of KwaZulu–Natal, Private Bag X54001, Durban 4000, South Africa

[2]Council for Scientific and Industrial Research, P.O. Box 395, Pretoria 0001, South Africa

*Corresponding author - email: aforbes1@csir.co.za , Tel: +27 12 841 2368, Fax: +27 12 841 3152

**It is well-known[1-3] how to control the spatial output from a laser, with most solutions to date involving customised intra-cavity elements in the form of apertures, diffractive optics and free-form mirrors. These optical elements require considerable design and fabrication effort and suffer from the further disadvantage of being immutably connected to the selection of a particular spatial mode. Consequently, most laser systems are designed for the ubiquitous Gaussian mode, whereas it is clear that there are many instances when a customised mode would be preferable. We overcome these limitations with the first digital laser, comprising an electrically addressed reflective phase-only spatial light modulator as an intra-cavity holographic mirror. The phase and amplitude of the holographic mirror may be controlled as simply as writing a new gray-scale image (computer generated hologram) to the device: on-demand laser modes. We show that we can digitally control the laser modes with ease, albeit with higher round-trip losses and thus requiring higher gain, and demonstrate the versatility of the technique by switching between several spatial modes in an otherwise standard solid-state laser resonator.**

Laser beam shaping tools have matured over the past few decades to allow external (to the laser cavity) reshaping of a laser beam to a desired transverse profile[4]. The procedures for determining the desired optical transformation are well known[4], and may be implemented by a variety of methods, for example, by diffractive optical elements, free-form optics or more recently by digital holograms written to a spatial light modulator (SLM). However, there are advantages to rather shaping the light *inside* the laser cavity (intra-cavity laser beam shaping) and this has been a subject of interest for a number of years[2,3], with several design techniques available[5-10], some of which have successfully been implemented, for example using phase-only[11-14], amplitude-only[1-3,15,16], and optically-addressed liquid crystal[17] optical elements for spatial mode selection. All of these techniques require custom optics, for example, a diffractive mirror or phase plate designed for a specific mode, while the optically-addressed liquid crystal approach requires external beam shaping (e.g., diffractive optics or SLMs) to address the optic, a wavefront sensor and optimization routine to iterate towards the desired phase profile, and thus results in unconventional, metre-long, cavities. There have also been attempts at dynamic intra-cavity beam control with deformable mirrors[18-23], but such elements have very limited stroke, are limited in the phase profiles that can be accommodated[18,19], and thus have found little application in laser mode shaping. Rather, such mirrors have been instrumental in high power applications such as correcting mode distortions (e.g., due to thermal load) or to maximizing energy extraction and optimization of laser brightness[20-23]. To date no technique has been demonstrated for the on-demand selection of laser modes.

Here we overcome the aforementioned limitations to intra-cavity laser beam shaping through the use of intra-cavity digital holograms, implemented on a phase-only reflective SLM, to form a rewritable holographic mirror in place of the standard laser cavity mirror. As we will show, this allows on-demand mode selection with high resolution and with a very wide dynamic range of phase values. Importantly, this "digital laser" may be used to implement amplitude-only, phase-only or amplitude and phase modulation by simply altering the digital hologram (gray-scale picture) written to the device.

Our laser cavity, as shown in **Figure 1**, consists of a conventional folded resonator configuration with a Nd:YAG laser crystal as the gain medium [see Methods]. What is unconventional is the use of a phase-only reflective SLM as the back optical element of the cavity, thus creating a digitally addressed holographic mirror. The key properties required of the SLM for this application are high resolution, high efficiency, high reflectivity at the desired polarisation, small phase-amplitude cross-talk, reasonable damage threshold and a large phase shift at the laser wavelength.

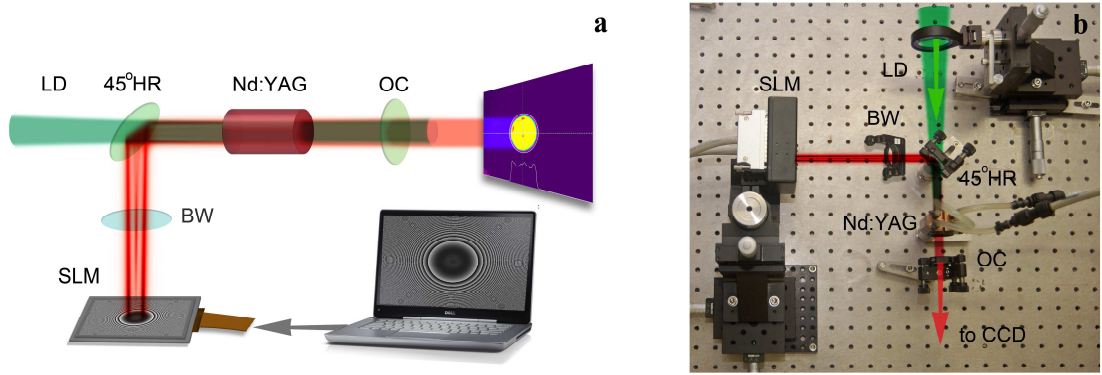

**Figure 1 | Concept and experimental realisation of the digital laser. a**, Schematic of the digital laser concept. **b**, Photograph of the experimental set-up.

The laser is optically pumped by a high-power diode laser that is coupled into the cavity through a mirror coated for high transmission at the diode wavelength (808 nm) and high reflectance at the lasing wavelength (1064 nm). This folding mirror forms an L-shaped cavity so that the high power diode beam does not interact with the SLM, thus avoiding damage. An important feature of the cavity is the intra-cavity Brewster window to force the laser to oscillate in the desired polarization for the SLM (vertical in our setup).

As a proof-of-principle experiment we programmed the holographic mirror to mimic a conventional concave mirror with a radius of curvature, $R$, chosen to ensure that the resonator formed a stable plano-concave cavity [**Figure 2 (a)**]. This requires a digital hologram of a lens to be programmed to the SLM, with focal length $f = R$, so that the hologram mimics the curvature of the mirror. The waist size (at the flat output coupler) of the Gaussian mode that oscillates in such a cavity may be described analytically as[1]

$$w_0^2 = (\lambda/\pi)[L(R-L)]^{1/2}, \tag{1}$$

where $L$ is the effective length of the resonator and $\lambda$ is the laser wavelength. Prior to testing the digital laser, two physical concave mirrors were used (separately) in the same set-up in place of the SLM, and the Gaussian beam size recorded at the output. The results for these two cases, $R = 400$ mm and $R = 500$ mm, are shown in **Figure 2 (b)** and plotted in **Figure 2 (c)** together with the theoretical curve following equation (1). The same test curvature examples were programmed digitally and are shown alongside the physical mirror measurements in **Figure 2 (b)**. From a mode selection perspective the laser performs identically in the two configurations. Next, as the digital hologram's programmed curvature was changed, **Figure 2 (c)**, so the measured output Gaussian beam size changed in accordance with equation (1). This confirms that the digital laser behaves as a standard stable cavity and it is clear from the results that the SLM mimics the stable cavity with high fidelity. In addition to confirming the desired behaviour of the digital laser, this experiment also brings to the fore another practical advantage: whereas with physical mirrors it is commonplace to have a limited and discrete selection on hand, with the digital approach virtually any mirror curvature can be created, on demand, by simply changing the gray-scale image representing the digital hologram, and is limited only by the resolution of the SLM used.

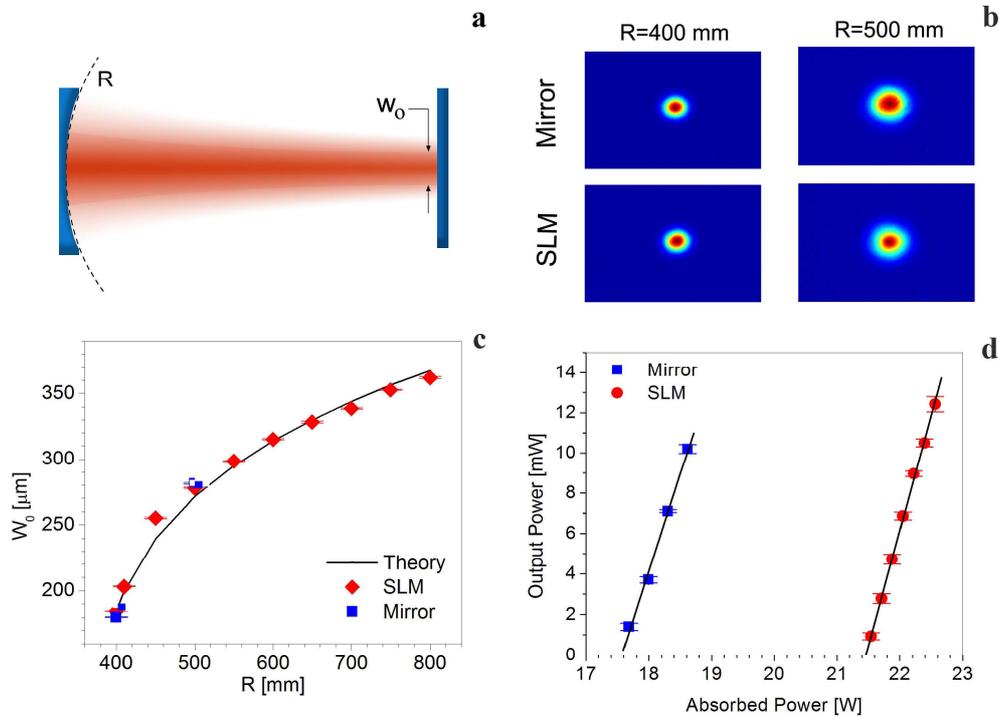

**Figure 2 | Comparison of the digital laser to an equivalent plano-concave laser. a**, Schematic of the stable plano-concave resonator with a waist plane at the flat output coupler. **b**, Measured intensity profiles for two curvature cases ($R = 400$ mm and $R = 500$ mm), comparing the digital laser output (SLM) to that of physical mirrors (Mirror). **c**, The change in measured beam size with digitally imposed curvature matches the theoretical curve. **d**, The threshold of the digital laser is higher than that of the conventional laser due to the additional losses from the SLM, shown here for the $R = 400$ mm case.

The higher losses of the SLM do manifest themselves as a higher threshold for lasing, as noted in **Figure 2 (d)**. Thus two conditions must be simultaneously satisfied for the digital laser to function: the gain of the laser must be sufficiently high to overcome the losses, but the intra-cavity circulating intensity must not exceed the damage threshold of the SLM. We manage this by virtue of a high-power pump source and an L-shaped cavity, but there are several other valid approaches (e.g., increasing the doping concentration of the crystal). When these conditions are balanced, the digital laser functions as designed.

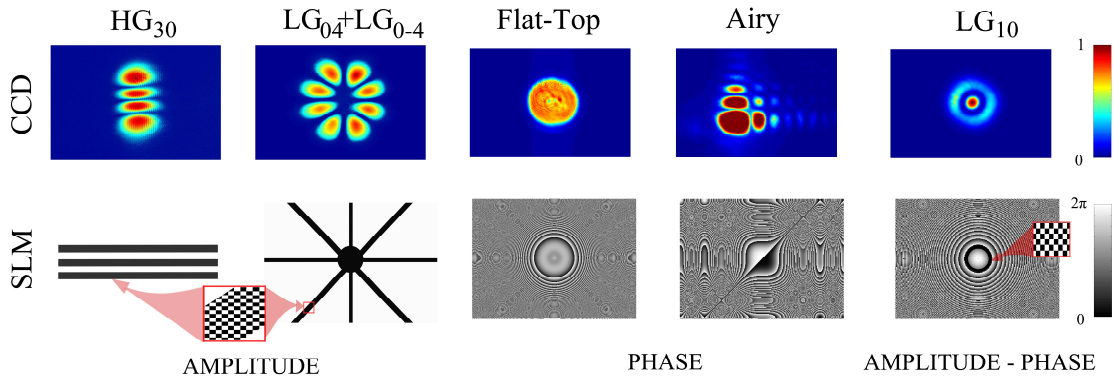

**Figure 3 | Customised spatial modes by amplitude and phase modulation.** By complex amplitude modulation a customised set of high-loss regions create a Hermite-Gaussian beam ($n = 3$, $m = 0$) and a superposition of Laguerre-Gaussian beams ($p = 0$, $l = \pm 4$) as the laser output. By phase-only modulation a flat-top beam and Airy beam are created as the stable modes of the cavity. Combining amplitude and phase effects allows for the selection of a Laguerre-Gaussian beam ($p = 1$, $l = 0$) of a chosen size.

As a final test we employ the digital laser to select the well-known Hermite-Gaussian, Laguerre-Gaussian, super-Gaussian (flat-top) and Airy beams. The selection of the Hermite-Gaussian and Laguerre-Gaussian modes [**Figure 3**] exploited complex amplitude modulation to implement amplitude modulation on the phase-only SLM[24]. In other words, the SLM can be used to create customised apertures, for example, the fine wires (loss-lines) used in the past for Hermite-Gaussian mode selection[1,2]. The digital hologram for the creation of the radial Laguerre-Gaussian beam ($p = 1$, $l = 0$) comprised of a high-loss annular aperture together with a phase-only radius of curvature. The former was set to the first zero of the Laguerre-Gaussian mode while the latter was used to select and control the mode size. Finally, many techniques exist for the design of intra-cavity diffractive optics[5-10] for particular mode selection, all of which may readily be applied to the digital laser. We illustrate this in **Figure 3** where an Airy beam[25] and flat-top beam[7] are created by phase-only digital holograms. Finally we note that the switching from one mode to another required nothing more than a change to the gray-scale image making up the digital hologram – no realignment and no additional optical elements. Traditionally, to create the spectrum of modes shown in **Figure 3** would require several laser resonator set-ups, each with a custom (expensive) optic.

In conclusion, we have demonstrated a novel digital laser that allows arbitrary intra-cavity laser beam shaping to be executed on the fly. We have shown that the digital laser can replicate conventional stable resonator cavities as well as "custom" laser resonators to produce more exotic laser modes. The digital laser is at present limited in the power that it can output, but this may be overcome with careful engineering of bespoke liquid crystals. Just as SLMs external to the laser cavity have proved an excellent means for testing high power beam shaping elements prior to fabrication, and have in the process opened up many avenues for low average power applications of structured light (e.g., holographic optical tweezers) so the digital laser may well become an robust, easy-to-implement, test bed for intra-cavity beam shaping ideas. Moreover, since the digital laser is rewritable one can envisage dynamic intra-cavity beam shaping, for example, in controlling thermal lensing and aberrations in real-time, to real-time mode control and switching. Customised laser modes are now only a picture away.

## Methods

Several spatial light modulators (SLMs) were used in the testing of the digital laser, and finally a Hamamatsu (LCOS-SLM X110468E) series device was selected for the digital laser. Previous tests with other SLMs failed mainly due to the phase-amplitude coupling which becomes pronounced during intra-cavity operation. The gain medium was a 1% doped Nd:YAG crystal rod with dimension of 30 mm (length) by 4 mm (diameter). The crystal was end-pumped with a 75 W Jenoptik (JOLD 75 CPXF 2P W) multimode fibre-coupled laser diode operating at 808 nm. The output coupler (flat curvature) had a reflectivity of 60% while the SLM had a measured reflectivity of 91% at the desired polarisation (vertical) and 93% at the undesired polarisation (horizontal). In order to force the cavity to lase on the vertical polarisation, an intra-cavity Brewster plate was used. On this polarisation, calibration tests on the SLM reveal typical efficiencies of ~86% into the first order and ~1% into the zeroth order. In the intra-cavity configuration this large difference results in suppression of the zeroth order due to the significantly higher round trip losses, and thus the SLM could be operated at normal incidence and without a grating. The SLM efficiency had a standard deviation of ~0.4% across all gray levels, i.e., minimal amplitude effects during phase modulation. The nominal length of the cavity was approximately 390 mm, but was determined to have an effective length of 373 mm in order to compensate for the small thermal lensing due to pump absorption in the crystal as well as the refractive index of the crystal. The effective length was used in all calculations for the mode sizes. The resonator output was 1:1 imaged onto a Spiricon CCD camera for intensity measurements, and could also be directed to a second SLM for modal decomposition studies. For far field tests the first lens of the telescope was removed.

# Figure captions

**Figure 1 | Concept and experimental realisation of the digital laser. a**, Schematic of the digital laser concept. **b**, Photograph of the experimental set-up.

**Figure 2 | Comparison of the digital laser to an equivalent plano-concave laser. a**, Schematic of the stable plano-concave resonator with a waist plane at the flat output coupler. **b**, Measured intensity profiles for two curvature cases ($R = 400$ mm and $R = 500$ mm), comparing the digital laser output (SLM) to that of physical mirrors (Mirror). **c**, The change in measured beam size with digitally imposed curvature matches the theoretical curve. **d**, The threshold of the digital laser is higher than that of the conventional laser due to the additional losses from the SLM, shown here for the $R = 400$ mm case.

**Figure 3 | Customised spatial modes by amplitude and phase modulation.** By complex amplitude modulation a customised set of high-loss regions create a Hermite-Gaussian beam ($n = 3$, $m = 0$) and a superposition of Laguerre-Gaussian beams ($p = 0$, $l = \pm 4$) as the laser output. By phase-only modulation a flat-top beam and Airy beam are created as the stable modes of the cavity. Combining amplitude and phase effects allows for the selection of a Laguerre-Gaussian beam ($p = 1$, $l = 0$) of a chosen size.